# How has the war in Ukraine affected Russian sentiments?

Mikael Elinder[1]†, Oscar Erixson[2]*†, Olle Hammar[3,4]†

**Affiliations:**

[1]Department of Economics, Uppsala University; Uppsala, Sweden.

[2]Institute for Housing and Urban Research, Uppsala University; Uppsala, Sweden.

[3]Department of Economics and Statistics, Linnaeus University; Växjö, Sweden.

[4]Institute for Futures Studies; Stockholm, Sweden.

*Corresponding author. Email: oscar.erixson@ibf.uu.se

†These authors contributed equally to this work.

**Abstract:** Using individual-level data from Gallup World Poll and the Levada Center, we provide an in-depth analysis of how Russia's invasion of Ukraine has affected sentiments in the Russian population. Our results show that after the invasion, a larger share of Russians expressed support for President Putin, optimism about the future, anti-West attitudes, and lower migration aspirations. The 2022 mobilization and the 2023 Wagner rebellion had only short-lived and no effects on sentiments, respectively. Regions with low pre-war support for Putin displayed stronger rally effects, higher casualty rates, and increased incomes, suggesting a recruitment strategy that maximizes political support. Taken together, our results suggest strong public support for the war, except among Russians abroad who became more critical of Putin in line with global views.

On February 24, 2022, Russia invaded Ukraine. Two years later, on May 7, 2024, Vladimir Putin was inaugurated as President for a fifth term with 88 percent of the reported votes and massive support in public opinion polls (*1–2*). Given that polls indicated record-low support for Putin before the invasion (*3*), the boost in his popularity suggests that the invasion spurred strong rally 'round the flag effects (*4*) in the Russian population (*5*). Similar boosts in Putin's popularity have been documented in connection with both the annexation of Crimea in 2014 (*6–8*) and the war in Georgia in 2008 (*9*), which suggests that Putin increases his popular support by military interventions in neighboring countries.

Relying on election results to assess Putin's popularity and Russians' sentiments about the war is however problematic, since election results are likely to be manipulated (*10–13*). Similarly, in authoritarian countries, fear of repercussions may lead citizens to withhold their true preferences in political opinion polls (*14–16*), although the extent of this problem in the Russian context is debated (*5, 17*).

Understanding the political dynamics of the war is crucial for identifying effective measures to end it. Currently, scholars and political experts hold differing views on the key drivers behind Putin's decision to invade Ukraine. Some argue that the war stems from Putin's personal ambition to restore the former Soviet empire's global power (*18*), while others argue that the historical narrative is used as justification for the invasion which serves Russia's contemporary geopolitical interests (*19*). Another argument is that the annexation of Crimea and the subsequent invasion of Ukraine are diversionary strategic moves by Putin to increase his popularity and strengthen his political power domestically (*5, 7*). Hence, an important question for understanding the motives behind the war is to what extent it is supported by the Russian population.

The war has undoubtedly been accompanied with negative consequences for many Russians. More than half a million young men have been sent to the frontlines, and by September 27, 2024, more than 71,000 of those were confirmed dead by Mediazona's recorded names count (*20*). The true number is likely much higher with hundreds of thousands killed or seriously injured. Moreover, domestic policies, ranging from restrictions on media and freedom of speech (*21*), to political oppression (*22*), along with increased military expenditures (*23*) and economic sanctions from the West (*24–26*), have dramatically deteriorated the opportunities of many Russians (*27*).

At the same time, there are reports indicating that life in Russia goes on as usual (*28–29*), and that many Russians believe Russia must defend itself against the West and the expansion of NATO (*7, 30*). Consequently, the war with Ukraine also appears to have spurred pride and nationalism among the Russian population (*30–31*).

In other words, the war can affect many different sentiments. Different segments of the Russian population may also have different feelings and attitudes about whether Russia has become a better or worse country to live in after the invasion, and if Putin is leading the country in the right direction or not.

The aim of this study is to analyze how the war in Ukraine has affected Russians' support for Putin as well as other sentiments and opinions. We do this by using individual-level microdata from two different and highly respected polling institutions: the Levada Center, an independent, non-governmental polling and sociological research organization in Moscow (*32*), and the Gallup World Poll (GWP), the most comprehensive and farthest-reaching survey of the world (*33*). GWP is conducted annually with between 2,000–4,000 respondents in Russia per year, and the Levada survey is conducted monthly with a sample size of approximately 2,000 respondents per month. Importantly, using two independent surveys—one from a Russian institution and one from an American institution—allows for cross-validation of the findings.

While Russians might be reluctant to truthfully report their opinions about Putin, we also analyze responses to other, less sensitive questions, such as their optimism about the future, current life satisfaction, attitudes toward the West, and desires to move abroad. As such, we can assess if changes in Putin's approval ratings are consistent with respondents' sentiments in other dimensions. Moreover, we use the unexpected timing of the invasion, as well as the partial military mobilization of young men, and the Wagner Group rebellion, to elicit plausibly causal effects of various features of the war on sentiments. In addition, we carefully examine how different segments of the Russian population responded to these events. This is important, as aggregate statistics may hide important heterogeneities in the population. To further understand the political dynamics of Russia's war policies, and the development of Russian sentiments during the war, we also analyze the regional differences in rally 'round the flag effects and their correlations with war casualties and income changes. Finally, we also use GWP to analyze how the war has impacted the approval of Putin among Russians abroad.

While the sample in both the GWP and Levada surveys are selected to be representative of the Russian population, estimates of rally effects could be biased if the people responding to the surveys after the invasion are systematically different compared to those who responded before the invasion. We assess this concern and show that the pre- versus post-invasion samples appear to be strongly balanced in terms of respondents' observable characteristics, that there seems to be no effect of the invasion on the share of respondents answering "Don't know" or "Hard to answer", and that the survey periods have been of similar lengths across the different waves (see fig. S1 in the Supplementary Materials). These tests increase the confidence that the results in this study reflect causal effects of the war on Russians' sentiments.

## Results

### *The invasion spurred positive sentiments in the Russian population*

Columns 1 and 2 in Fig. 1 display time-series data on sentiments in the Russian population from the Levada (column 1) and GWP (column 2) surveys, before and after the invasion. Both surveys show that support for Putin (**A**) increased following the invasion and remained on a higher level during the first year of the war. The immediate effect shows an increase of 13 percentage points between February and March in Levada, and the more long-run effect an increase of 25 percentage points between 2021 and 2022 in GWP.

Regarding the other sentiments (**B**–**E**), both surveys reveal a similar persistent increase in optimism about the future (**B**), a decrease in the share with a positive attitude towards the West (**D**), and a reduction in the share of Russians who would like to move abroad (**E**). The response in subjective well-being (**C**), however, is less consistent; while Levada does not reveal any change in current mood following the invasion, GWP suggests a surge in life satisfaction.

In column 3, we provide estimates of the immediate effects of the invasion on sentiments, as measured by a comparison of responses to the Levada survey conducted just before (February 14–20) versus just after (March 27–April 2) the invasion. The effects, except for mood, are all statistically significant at the 5 percent level. Estimates based on the GWP data from 2021 versus 2022 (bottom estimates in column 3) confirm the Levada estimates, but also reveal a statistically significant increase in life satisfaction following the invasion.

Heterogeneity analyses, based on the monthly data from Levada (column 3), indicate that the immediate responses are remarkably similar across various demographics and population groups, consistent with broad popular support for the invasion of Ukraine. The small differences in point estimates are generally not statistically different from the average effect. The main exception is found for residents in Moscow, who respond less to the invasion, with statistically insignificant effects.

Taken together, the analysis indicates that the invasion had large and persistent impacts on a variety of sentiments in broad segments of the Russian population. These results are also confirmed in a large set of robustness and sensitivity analyses (see table S11 in the Supplementary Materials).

[Fig. 1]

***The partial military mobilization created just a temporary crack in the war support***

On September 21, 2022, Vladimir Putin declared a partial military mobilization of recruits for the war. One month later, on October 28, the mobilization was announced completed. During the mobilization, all men of conscription age, 18–27 years old, were at risk of being sent to the frontlines in Ukraine. While young men were thus most likely to be directly affected by the mobilization, other population groups, such as parents and partners, may also have reacted to this policy.

Fig. 2 shows Russians' sentiments during the months surrounding the mobilization. As shown in column 1 (A–D, based on Levada), the immediate impact of the mobilization was a clear negative effect on the support for Putin (**A**), optimism about the future (**B**), current mood (**C**), and support for the military actions of the Russian Armed Forces in Ukraine (**D**). These effects, however, were short-lived and had all disappeared within one to five months.

If we look at the heterogeneity of these immediate effects (column 3), we can see that young men, who were more directly targeted by the mobilization, did not seem to respond more strongly than other groups. Young women, however, appeared to respond more negatively in terms of their optimism about the future, and old women more negatively in terms of their current mood. A breakdown of the respondents' mood shows that there was an increase in the feelings of fear, melancholy, tension and irritation during the month of the mobilization, while no such mood changes were shown at the time of the invasion (see fig. S2 in the Supplementary Materials).

In GWP (column 2), the 2022 polling period in Russia was between August and November, hence coinciding with the mobilization. A comparison of responses among those interviewed before (August 13–September 20, $n = 942$) versus after (September 21–November 2, $n = 1,064$) the announcement shows a negative and marginally statistically significant effect on subjective well-being (**C**). However, the other sentiments were largely unaffected, which confirms the mobilization's short-lived impact. For migration aspirations (**E**), we note a positive, but not statistically significant, increase in young men's willingness to leave the country.

In sum, this analysis suggests that while the mobilization appears to have been broadly disliked when announced, this effect only lasted until its completion. As such, it essentially just created a temporary crack in the generally positive view of the war.

[Fig. 2]

### *Russians appeared indifferent to the Wagner Group rebellion*

Another significant war event that may have impacted Russian sentiments and the support for Putin was the Wagner Group rebellion led by Yevgeny Prigozhin on June 23, 2023. Although the rebellion only lasted for one day, there were reports of people supporting the Wagner troops as they captured the Russian city Rostov-on-Don and advanced towards Moscow (*34*).

By exploiting the fact that the rebellion took place in the middle of the 2023 GWP polling period in Russia, we can investigate whether it had any significant impact on sentiments. A comparison of responses in the weeks just before (May 23–June 22, $n = 893$) versus just after (June 23–July 29, $n = 1,124$) the rebellion shows that it had no statistically significant effect, neither positive nor negative, on the support for Putin or other sentiments (see Fig. 3).

[Fig. 3]

### *Strategic recruitment, war casualties, and economic compensations*

According to the rally 'round the flag theory, there should be more room for rally effects when political support is low (*35*). Consistent with this, we find that in federal districts where the support for Putin was lower before the invasion, the rally effects of the invasion were stronger (see Fig. 4, **A**). Since stronger rally effects can also be interpreted as stronger support for the war, recruiting soldiers from regions with stronger rally effects might be less costly in terms of political support. Analyzing Mediazona's data on Russian casualties in Ukraine (*20*), as a proxy for recruitment intensity, we find that the number of confirmed military deaths per capita indeed is higher in regions with stronger rally effects from the invasion (**B**), suggesting that recruitment has been more intense in regions with stronger support for the war. However, as casualties accumulate, the political support for war in foreign countries is expected to decrease (*36–37*). In contrast to this prediction, we do not find any evidence that the support for Putin fades more quickly in regions with more casualties (**D**). While this can partly be explained by a stronger support for the war in these regions, another contributing factor could be that that soldiers' families gain financially from the war, as the government pays out economic compensations to families in which a member has been injured or killed (*38*). In line with this explanation, we find that incomes have grown more rapidly during the war in regions with more casualties (**C**). Taken together, these results are consistent with a strategic war plan that maximizes the political support for Putin.

[Fig. 4]

*Russians abroad have turned against Putin*

Our analyses above show that the Russian population in general seem to be supportive of Putin and the war in Ukraine. But how is the war perceived among Russians living outside of Russia?

To analyze this, we use a question in GWP about the approval of Russia's leadership, which has been asked annually in more than 100 countries all around the world. Regarding the invasion, there is a sharp drop in approval rates between 2021 and 2022 in these countries, both in anti-Putin and pro-Putin ones, suggesting a global dislike of the war (Fig. 5). In fact, there are only six countries outside of Russia (Afghanistan, Algeria, Bangladesh, India, Mali, and Tunisia) in which the invasion had a positive effect on the approval of Putin (see fig. S3 in the Supplementary Materials). It can also be noted that, while the annexation of Crimea in 2014 and the war in Georgia in 2008 both increased the support for Putin in Russia with rally effects of similar magnitudes (29 and 24 percentage points, respectively, see table S9 in the Supplementary Materials), these events were not associated with as large drops in the foreign approval rate of Putin as the 2022 full-scale invasion (**A**).

Analyzing the approval of Putin among Russians abroad (that is, people born in Russia or with a Russian nationality), we find a negative effect of the invasion, suggesting that, unlike their countrymen in Russia, Russians abroad generally oppose the war. (**B**) shows that this effect is particularly pronounced among Russians in anti-Putin countries, where it is even stronger than for the general population. We can also note that, in most previous years, the support for Putin has been higher among Russians abroad than those living in Russia. With the 2022 invasion of Ukraine, however, Putin's support among Russians abroad has diverged from the population in Russia and instead converged with the worldview outside of Russia. In other words, the domestic rally effects have this time come at the cost of the previously patriotic diaspora's support, who now for the first time ever mostly disapprove of Putin.

[Fig. 5]

**Concluding remarks**

Our analysis, based on two independent surveys, shows strong and persistent rally 'round the flag effects, in broad segments of the Russian population, on a wide range of sentiments. These results indicate strong support for the invasion of Ukraine among Russians, suggesting that the war is unlikely to be ended due to public uprising within a foreseeable future.

It should be noted, however, that our analysis does not explain why we observe such strong rally effects following the invasion. One plausible explanation is the Kremlin's influence on the narrative in Russian media, along with direct censorship and propaganda (*39–40*). Western sanctions against Russia have also likely contributed to the polarization between Russia and the West, including the observed rise in anti-Western sentiments within Russia.

At the same time, our analysis also indicates that Russians appeared to dislike the partial mobilization, which may be one reason why Putin has postponed a larger general mobilization.

Moreover, as casualties accumulate, the compensation scheme for war casualties may eventually become too costly for the Russian state budget, which may then weaken public support for the war, especially in the most severely affected regions. Such an outcome, however, is closely related to the extent, enforcement, and efficacy of the economic sanctions against the country (*26*).

Finally, our analysis indicates that the Russian diaspora, which previously has been supportive of the Russian leadership, has now turned against Putin, in accordance with the rest of the world. Although Putin appears to care little about the outside world's view of Russia, it is possible that the sentiments among Russians abroad may eventually spread to relatives and friends in Russia.

**Acknowledgments:** We thank Konstantin Sonin and seminar participants at the Annual Congress of the IIPF, ASWEDE Conference, ifo Workshop on Return and Integration Prospects of Ukrainian Refugees, Linnaeus University, Migration Studies Delegation (Delmi), Research Institute of Industrial Economics (IFN), Stockholm Institute of Transition Economics, Uppsala Immigration Lab and Urban Lab for their valuable comments. We thank Delmi for providing access to the Gallup World Poll data.

**Funding:**

Jan Wallanders och Tom Hedelius stiftelse samt Tore Browaldhs stiftelse grant Fh22-0044 (ME, OE)

Johan och Jakob Söderbergs stiftelse grant FA22-0025 (OH)

Swedish Research Council for Health, Working Life and Welfare (Forte) grant 2022-00859 (OH)

**Author contributions:**

Conceptualization: ME, OE, OH

Methodology: ME, OE, OH

Investigation: ME, OE, OH

Visualization: ME, OE, OH

Funding acquisition: ME, OE, OH

Project administration: ME, OE, OH

Supervision: ME, OE, OH

Writing – original draft: ME, OE, OH

Writing – review & editing: ME, OE, OH

**Competing interests:** Authors declare that they have no competing interests.

**Data and materials availability:** The analysis is based on survey data from the Gallup World Poll (GWP) and the Levada Center. According to our agreements with these organizations, we are not permitted to share these data directly. However, the original microdata can be obtained through the same process as we did. Requests for getting access to the GWP data should be directed to galluphelp@gallup.com or at https://www.gallup.com/270188/contact-us-general.aspx, and for the Levada data to direct@levada.ru or at https://www.levada.ru/en/contact-us/. Data on Russian casualties by region were web scraped from Mediazona at https://en.zona.media/article/2022/05/20/casualties_eng. All codes used for the construction and analysis of the data will be made available in a replication package deposited in a public database after publication.

**Fig. 1. Effects of the invasion on sentiments in Russia.**

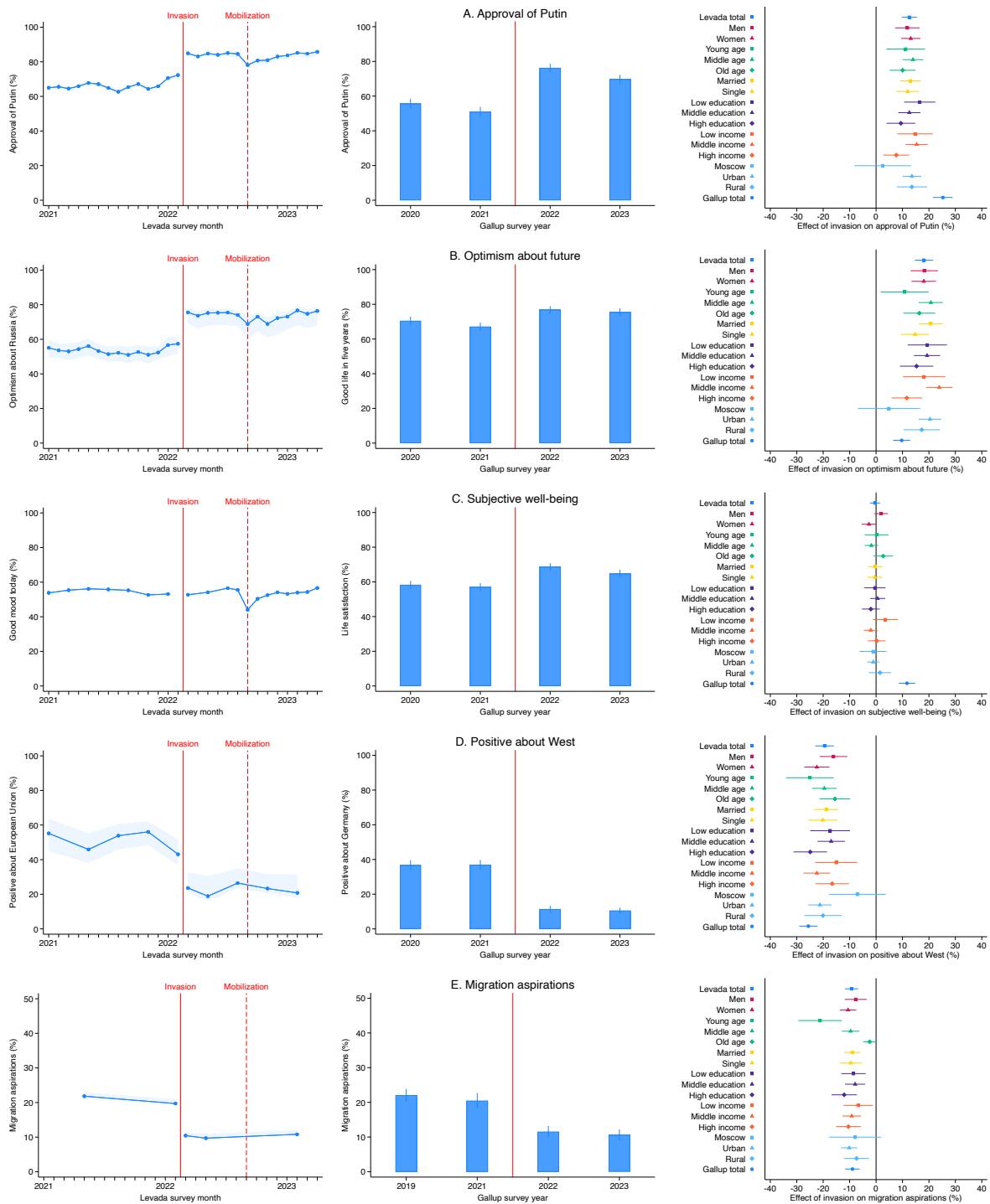

*Note:* The figure shows the effects of the Russian invasion of Ukraine on February 24, 2022, on the approval of Putin (**A**), optimism about the future (**B**), subjective well-being (**C**), attitudes about the West (**D**), and migration aspirations (**E**). Column 1 shows monthly averages from the Levada Center's public opinion surveys in Russia between January 2021 and April 2023, where the shaded areas show the results when all

"Hard to answer" responses are recorded as positive (upper bound) and negative (lower bound) responses. Solid red line indicates the full-scale invasion on February 24, and dashed line the mobilization on September 21, 2022. Column 2 shows yearly averages from the Gallup World Poll (GWP) in Russia between 2020 and 2023 (for migration aspirations, the first year is 2019 because the migration question was not asked in 2020), with 95 percent confidence intervals. Column 3 shows the invasion effects for various segments of the Russian population, estimated as the difference between the March (Mar. 27–Apr. 2) and February (Feb. 14–20) 2022 Levada surveys for each subgroup (for current mood the comparison is between January and March because the mood question was not asked in February 2022), with 95 percent confidence intervals estimated by a linear regression with robust standard errors. "Gallup total" estimates the difference between the 2022 (Aug. 13–Nov. 2) and 2021 (May 14–Jul. 14) survey waves in Gallup. For exact variable definitions, estimation methods, survey questions, sample sizes, and survey dates, see the Supplementary Materials.

**Fig. 2. Effects of the mobilization on sentiments in Russia.**

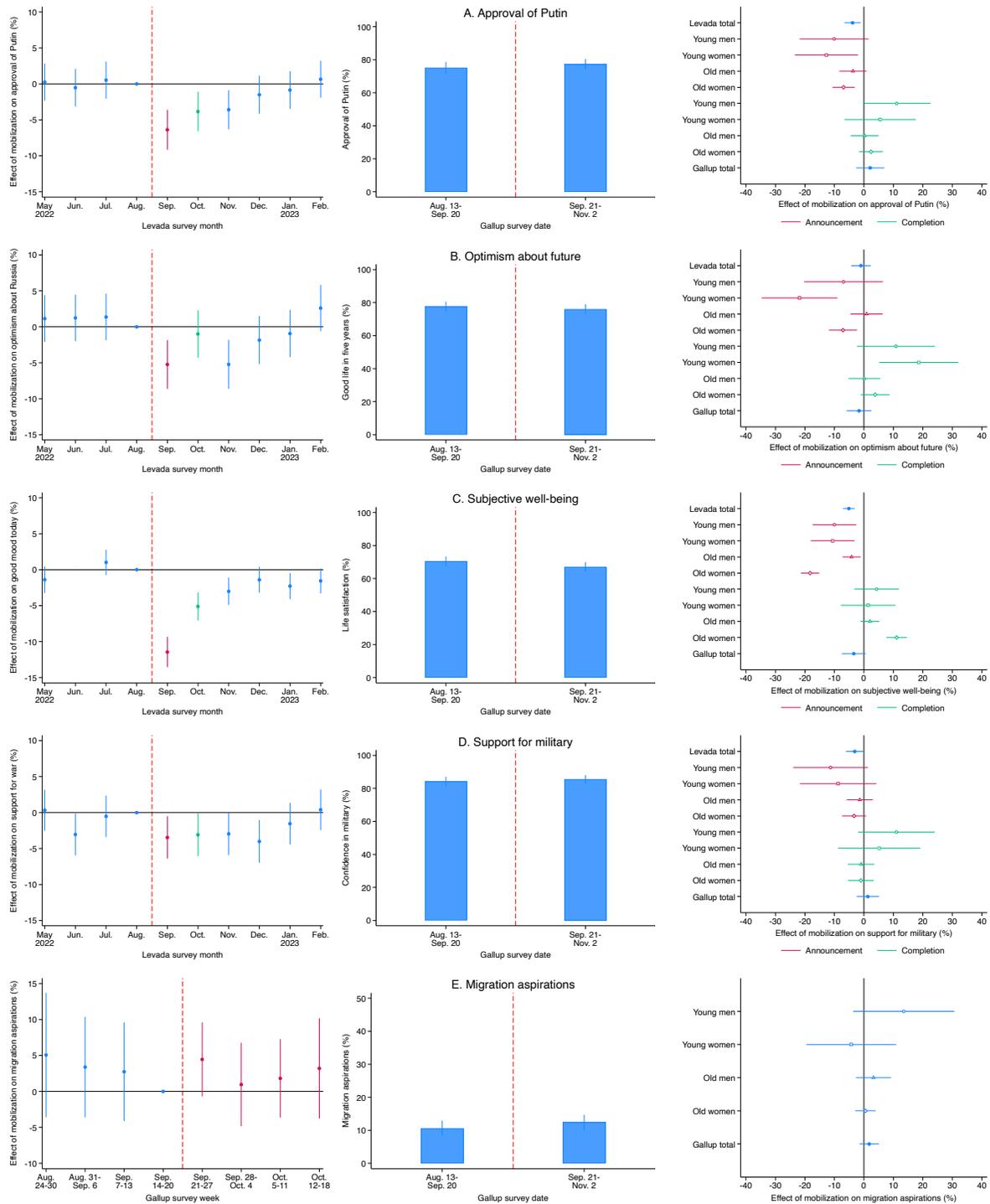

*Note:* The figure shows the effects of the partial military mobilization of young men (aged 18–27) that took place between September 21 and October 28, 2022, on the approval of Putin (A), optimism about the future (B), subjective well-being (C), support for the military (D), and migration aspirations (E). Column 1 shows the effects of the mobilization per month in Levada, estimated as the differences relative to August (Aug.

27–Sep. 2) 2022 with 95 percent confidence intervals and robust standard errors. Dashed line indicates the mobilization on September 21, 2022. Column 2 shows averages before and after the mobilization from the Gallup survey wave in 2022, with 95 percent confidence intervals. Column 3 shows the mobilization effects for young (ages 18–27) and old (ages 28–99) men and women in the Levada surveys, estimated as the difference between September (Sep. 24–30) and August 2022 for the announcement effect (red), and between October (Oct. 23–29) and September 2022 for the completion effect (green), with 95 percent confidence intervals and robust standard errors. "Levada total" estimates the total mobilization effect as the difference between the October and August 2022 Levada surveys. "Gallup total" estimates the mobilization effect in the 2022 Gallup survey wave. For migration aspirations, all columns are based on Gallup because the migration question was not asked in Levada in these months, and column 1 shows the effects of the mobilization per week. For exact variable definitions, estimation methods, survey questions, sample sizes, and survey dates, see the Supplementary Materials.

**Fig. 3. Effects of the Wagner Group rebellion on sentiments in Russia.**

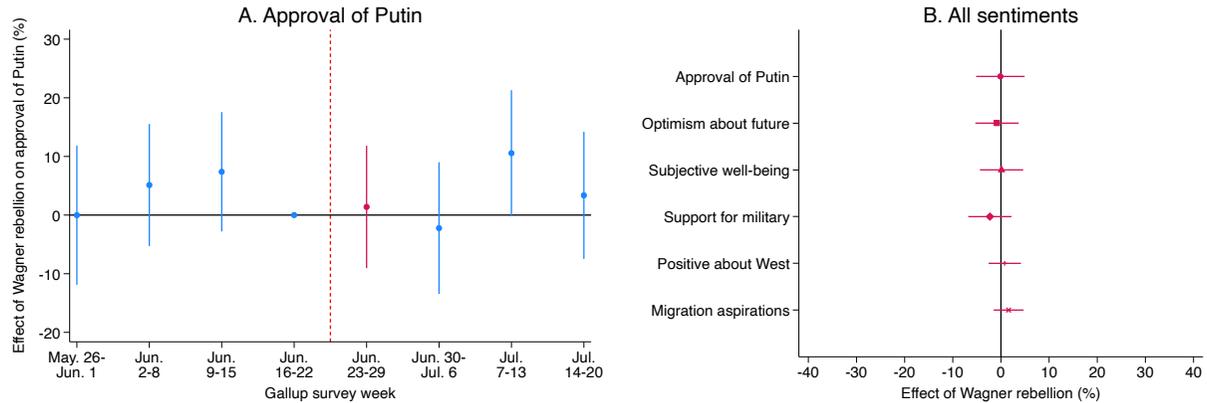

*Note:* The figure shows the effects of the Wagner Group rebellion on June 23–24, 2023, on the approval of Putin (**A**), optimism about the future, subjective well-being, support for the military, attitudes about the West, and migration aspirations (**B**). (A) shows the effects of the rebellion per week in Gallup, estimated as the differences relative to the week before the rebellion (Jun. 16–22), with 95 percent confidence intervals and robust standard errors. Dashed line indicates the rebellion on June 23, 2023. (B) shows the rebellion effects in the Gallup 2023 survey, estimated as the difference before (May 23–Jun. 22) and after (Jun. 23–Jul. 29) the rebellion, with 95 percent confidence intervals and robust standard errors. For exact variable definitions, estimation methods, survey questions, sample sizes, and survey dates, see the Supplementary Materials.

**Fig. 4. Regional casualties and support for Putin.**

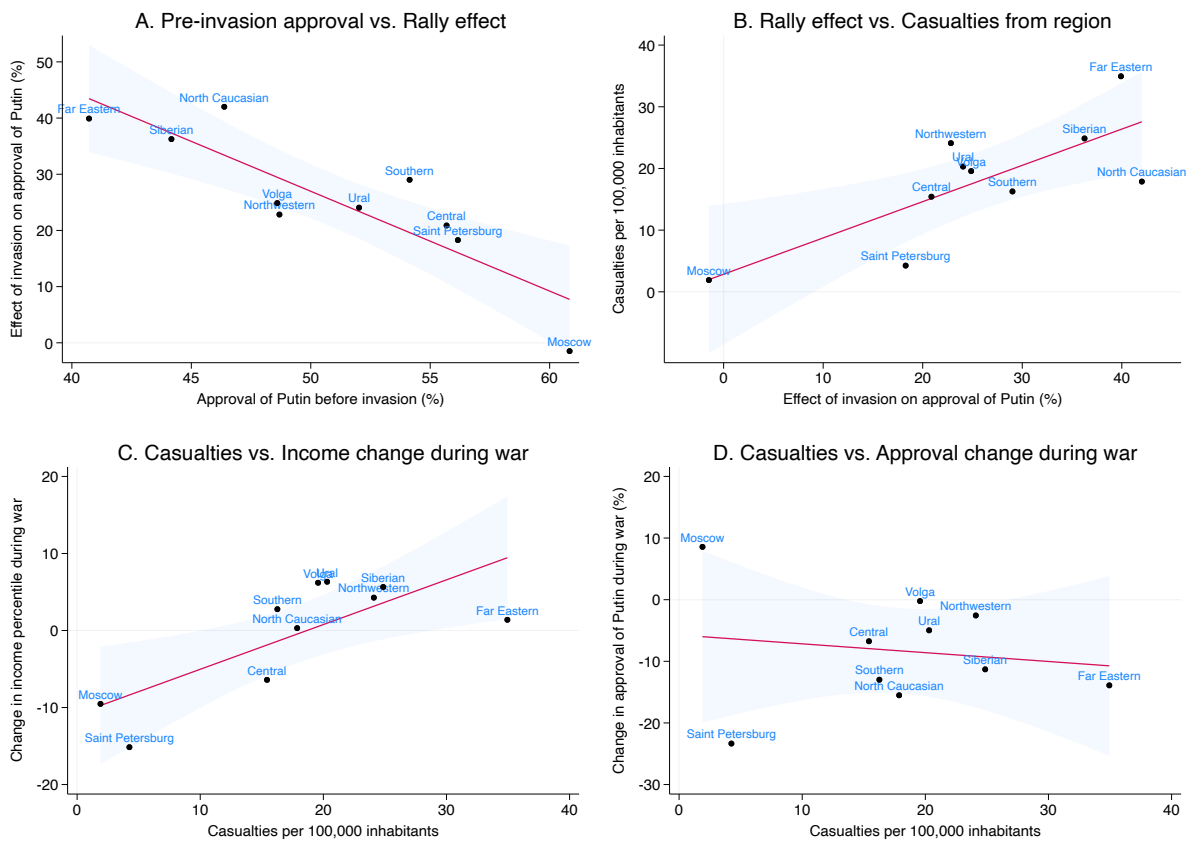

*Note:* (**A**) shows that the invasion rally effect on approval of Putin was higher in regions with lower pre-invasion support for Putin (Pearson's correlation coefficient, $r = -0.87$, and $p < 0.01$). (**B**) shows that the number of casualties per capita has been higher in regions with higher invasion rally effects ($r = 0.77$, $p < 0.01$). (**C**) shows that the relative income increase during the war has been higher in regions with higher casualties per capita ($r = 0.75$, $p < 0.05$). (**D**) shows no statistically significant relationship between regional-level casualties per capita and change in approval of Putin during the war ($r = -0.15$, $p > 0.1$). Red lines show linear predictions with 95 percent confidence intervals. Central region excludes Moscow, and Northwestern region excludes Saint Petersburg. Approval of Putin before invasion measured as average approval of Putin per federal district, Moscow and Saint Petersburg in the 2021 (May 14–Jul. 14) Gallup survey wave. Effect of invasion on approval of Putin estimated as the difference between the 2022 (Aug. 13–Nov. 2) and 2021 survey waves in Gallup for each region. Casualties per 100,000 inhabitants measured as the number of confirmed military deaths per region between February 24, 2022, and April 30, 2023, as reported by Mediazona and divided by population figures from the Federal State Statistics Service (Rosstat) in 2020. Change in income percentile during war measured as the difference in mean per capita income percentile between the 2023 (May 23–Jul. 29) and 2022 survey waves in Gallup for each region. Change in approval of Putin during war measured as the difference between the 2023 and 2022 survey waves in Gallup for each region. For exact variable definitions, estimation methods, survey questions, sample sizes, and survey dates, see the Supplementary Materials.

**Fig. 5. Long-run trends and support for Putin among Russians abroad.**

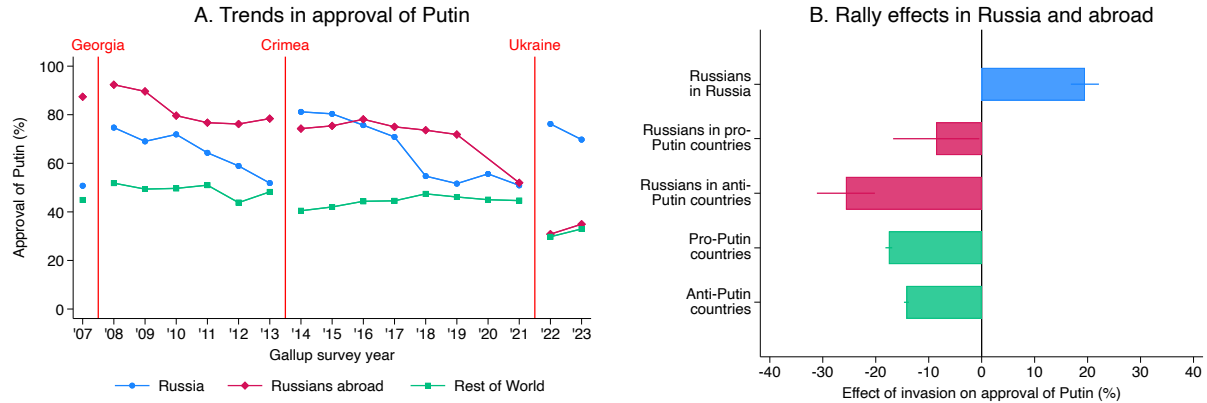

*Note:* The figure shows the approval of Putin in Russia (blue), among Russians abroad (red), and in the rest of the world (green). Russians abroad include people in other countries who were born in Russia and/or have a Russian nationality. (**A**) shows the trends in Gallup between 2007 and 2023, where solid lines indicate the war in Georgia in August 2008, the annexation of Crimea in February–March 2014, and the invasion of Ukraine in February 2022 (for Russians abroad, 2021 includes both 2020 and 2021 because of a small number of respondents for these questions in 2020 due to the COVID-19 pandemic). (**B**) shows the effects of the invasion, estimated as the difference between the 2022–2023 and 2020–2021 Gallup survey waves, with 95 percent confidence intervals and robust standard errors. Pro-Putin countries are defined as countries in which 50 percent or more approved of Putin in 2020–2021, and anti-Putin countries as countries where less than 50 percent of the population approved of Putin in 2020–2021. For exact variable definitions, estimation methods, survey questions, sample sizes, and survey dates, see the Supplementary Materials.

# Supplementary Materials for

## How has the war in Ukraine affected Russian sentiments?

Mikael Elinder, Oscar Erixson, Olle Hammar

Corresponding author: oscar.erixson@ibf.uu.se

**The PDF file includes:**

**Materials and Methods**

<u>Materials</u>

In this section, we describe the data used in the empirical analysis. We begin with general descriptions of the data sources and then provide a more detailed description of the variables used to construct the figures in the main text. Three different data sources are used for the analysis: 1) Gallup World Poll, 2) Levada Center, and 3) Mediazona.

***Data Sources***

*Gallup World Poll*

The Gallup World Poll (GWP) conducts annual surveys on attitudes and behaviors in more than 160 countries around the world, corresponding to 99 percent of the world's adult population (*33*). The survey includes at least 1,000 individuals per country and year, but in some large countries, including Russia, sample sizes of at least 2,000 individuals per year are collected. Gallup uses either telephone surveys, using a random-digit-dial method or a nationally representative list of phone numbers, or face-to-face interviews in randomly selected households using an area frame design. Face-to-face interviews are approximately one hour, and telephone interviews are about 30 minutes. The samples are probability based and nationally representative of the resident population aged 15 years and older. The coverage area is the entire country including rural areas, and the sampling frame represents the entire civilian, non-institutionalized adult population of the country. The final GWP samples are weighted to correct for unequal selection probability, nonresponse, and double coverage of landline and cellphone users when using both cellphone and landline frames. Gallup also weights its final samples to match the national demographics of each selected country.

In Russia, the mode of interviewing was face-to face in 2006–2019, landline and mobile telephone in 2020–2021, face-to-face in 2022, and mobile telephone in 2023. The interview language was Russian. In some years, people living in very remote or difficult to access areas were excluded. In these cases, the excluded areas represent five percent or less of the population (*41*).

In the analysis, we use the GWP data from Russia in 2007–2023. Russians abroad are defined as individuals who live in another country than Russia, but who were born in Russia and/or have Russian nationality. Rest of the world is the non-Russian GWP sample. The annual sample sizes and survey dates in GWP are shown in table S1.

*Levada Center*

The Levada Center is an independent, non-governmental polling and research organization based in Moscow (*32*). The center has conducted regular, nationally representative surveys and public opinion polls across Russia since 1988. Since 2016, it has been labelled a foreign agent under the Russian foreign agent law. The Levada data have been widely used in research (*5, 9, 17*) and are often considered the most reputable series of public opinion data in Russia (*5, 17*).

In the analysis, we use monthly, individual-level microdata from the Levada Center between January 2021 and April 2023. Data are weighted using the main vector. The Levada survey dates and sample sizes are shown in table S2.

*Mediazona*

For the number of casualties per region, we are using the Mediazona count on Russian losses in the war with Ukraine (*20*). In collaboration with BBC News Russian service and a team of volunteers, Mediazona maintains a named list of deceased Russian military personnel. This list is compiled from verified, publicly available sources, including social media posts by family members, local news reports, and official announcements from regional authorities. The list is not exhaustive, as not every military death becomes public knowledge.

In the analysis, we use Mediazona's number of confirmed military deaths (all troops) between February 24, 2022, and April 30, 2023, by Russian region in which they lived. The absolute numbers are adjusted for population by dividing them with regional populations in 2020 from the Federal State Statistics Service (Rosstat).

### *Variable Definitions*
*Approval of Putin*
Levada: "Do you generally approve or disapprove of the activities of the President of Russia?"

| | |
|---|---|
| 1 | Approve |
| 0 | Disapprove |
| [. | Hard to answer] |

GWP: "Do you approve or disapprove of the job performance of the leadership of this country?"

| | |
|---|---|
| 1 | Approve |
| 0 | Disapprove |
| [. | DK / Refused] |

*Optimism about Future*
Levada: "Do you think that things in the country are going in the right direction, or do you think the country is going the wrong way?"

| | |
|---|---|
| 1 | Things are going in the right direction |
| 0 | The country is going the wrong way |
| [. | Hard to answer] |

GWP: "Please imagine a ladder with steps numbered from 0 at the bottom to 10 at the top. Suppose we say that the top of the ladder represents the best possible life for you, and the bottom of the ladder represent the worst possible life for you. Just your best guess, on which steps do you think you will stand on in the future, say about five years from now?"

| | |
|---|---|
| 1 | Best possible (6–10) |
| 0.5 | 5 |
| 0 | Worst possible (0–4) |
| [. | DK / Refused] |

*Subjective Well-Being*
Levada: "What can you say about your mood in recent days?"

| | |
|---|---|
| 1 | In a great mode |
| 0.5 | Normal, even mood |
| 0 | I feel fear, melancholy / I feel tension, irritation |
| [. | Hard to answer] |

GWP: "Please imagine a ladder with steps numbered from 0 at the bottom to 10 at the top. Suppose we say that the top of the ladder represents the best possible life for you, and the bottom of the ladder represents the worst possible life for you. On which step of the ladder would you say personally feel you stand at this time, assuming that the higher the step the better you feel about your life, and the lower the step the worse you feel about it? Which step comes closest to the way you feel?"

| 1  | Best possible (6–10) |
|----|----------------------|
| 0.5 | 5 |
| 0  | Worst possible (0–4) |
| [. | DK / Refused] |

*Positive about West*
Levada: "How do you generally feel about the European Union now?"

| 1 | Very good / Mostly good |
|---|-------------------------|
| 0 | Very bad / Mostly bad |
| [. | Hard to answer] |

GWP: "Do you approve or disapprove of the job performance of the leadership of Germany?"

| 1 | Approve |
|---|---------|
| 0 | Disapprove |
| [. | DK / Refused] |

*Migration Aspirations*
Levada: "Would you like to move abroad for permanent residence?"

| 1 | Definitely yes / More likely yes |
|---|----------------------------------|
| 0 | Definitely no / More likely no |
| [. | Hard to answer] |

GWP: "Ideally, if you had the opportunity, would you like to move permanently to another country, or would you prefer to continue living in this country?"

| 1 | Like to move to another country |
|---|----------------------------------|
| 0 | Like to continue living in this country |
| [. | DK / Refused] |

*Support for Military*
Levada: "Do you personally support or not the actions of the Russian Armed Forces in Ukraine?"

| 1 | Definitely yes / More likely yes |
|---|----------------------------------|
| 0 | Definitely no / More likely no |
| [. | Hard to answer] |

GWP: "In this country, do you have confidence in each of the following, or not? How about the military?"

| 1 | Yes |
|---|-----|
| 0 | No |
| [. | DK / Refused] |

*Gender*
Levada / GWP: "Gender"

| Men | Male |
|-----|------|
| Women | Female |

*Age*
Levada: "Age"

| Young | 18–27 |
|-------|-------|
| Middle | 28–59 |

Old        60–99
GWP: "Please tell me your age."
    Young    18–27
    Old      28–99+
    [.       15–17 / Refused]

*Marital Status*
Levada: "Marital status"
    Married    Married / Not registered, but live together
    Single     Not registered, live separately / Single (not married), never been married / Live
               separately, but not divorced / Divorced / Widower (widow)
    [.         Refusal to answer]

*Education*
Levada: "Education"
    Low        Other
    Middle     Professional
    High       Higher education

*Income*
Levada: "How would you describe the material status of your family?"
    Low        We barely make ends meet, we don't even have enough money for food / We have
               enough money for groceries, but buying clothes causes financial difficulties
    Middle     We have enough money for groceries and clothes, but buying durable goods causes
               financial difficulties
    High       We can afford quite expensive household items / We can easily buy durable goods
    [.         Refusal to answer]

*Geographical Area*
Levada: "Size of the populated area"
    Moscow    Moscow
    Urban     Cities up to 100 thousand / From 100 to 500 thousand / More than 500 thousand
    Rural     Village

*Region*
GWP: "Region 3 Russia"
    Moscow           Moscow city (capital)
    Saint Petersburg Saint-Petersburg city
    Central          Center (excl Moscow city)
    Northwestern     Northwest (excl Saint-Petersburg city)
    Southern         South
    Volga            Privolzhskiy
    Ural             Urals
    Siberian         Siberia
    Far Eastern      Far East
    North Caucasian  North Caucasus

[.               DK / Refused]

*Casualties per 100,000 Inhabitants*
Mediazona: Casualties from region (Feb. 24, 2022–Apr. 30, 2023) / Population in region before invasion (2020) * 100,000

*Income Percentile*
GWP: "Per Capital income quintiles" * 20

| | |
|---|---|
| 20 | Poorest 20% |
| 40 | Second 20% |
| 60 | Middle 20% |
| 80 | Fourth 20% |
| 100 | Richest 20% |

*Russians Abroad*
GWP: "What is your nationality?" / "In which country were you born? (asked only of those who were not born in this country)"
      Russian / Russia

*Approval of Putin (Russians Abroad and Rest of World)*
GWP: "Do you approve or disapprove of the job performance of the leadership of Russia?"

| | |
|---|---|
| 1 | Approve |
| 0 | Disapprove |
| [. | DK / Refused] |

*Pro- and Anti-Putin Countries*
GWP: Mean approval of Putin in country

| | |
|---|---|
| Pro-Putin | Mean approval of Putin before invasion (2020–2021) 50–100% |
| Anti-Putin | Mean approval of Putin before invasion (2020–2021) 0–50% |

**Events**
*Invasion*
Russian invasion of Ukraine: February 24, 2022

| | Before | After |
|---|---|---|
| Levada | February (Feb. 14–20) | March (Mar. 27–Apr. 2) |
| GWP | 2021 (May 14–Jul. 14) | 2022 (Aug. 13–Nov. 2) |

*Mobilization*
Russian mobilization: September 21–October 28, 2022

| | Before | Announcement | Completion |
|---|---|---|---|
| Levada | August (Aug. 27–Sep. 2) | September (Sep. 24–30) | October (Oct. 23–29) |
| GWP | 2022 (Aug. 13–Sep. 20) | 2022 (Sep. 21–Nov. 2) | |

*Wagner Rebellion*
Wagner Group rebellion: June 23–24, 2023

| | Before | After |
|---|---|---|
| | Before | After |

| GWP | 2023 (May 23–Jun. 22) | 2023 (Jun. 23–Jul. 29) |

## Methods

In this section, we describe the methods used for the empirical analyses. Figures referred to below refer to the figures in the main paper.

### Methods for Fig. 1 in Main Paper

Column 1: Time-series graphs showing the average values in the monthly Levada data between January 2021 and April 2023. Mean values weighted using the main vector sampling weights. Solid lines show mean values for the full sample excluding "Hard to answer" responses. Shaded areas show results when all "Hard to answer" responses are recorded as either the maximum (upper bound) or the minimum (lower bound) values.

Column 2: Bars showing average values in the annual GWP survey waves between 2020 and 2023, with 95 percent confidence intervals. Mean values weighted using the GWP sampling weights. "Don't know" ("DK") and "Refused" responses excluded.

Column 3: Coefficient plots showing the effects of the invasion estimated using the following linear regression:

$$y_i = \alpha + \beta Post_i + \varepsilon_i$$

where $y_i$ is the survey response of individual $i$ for the relevant sentiment; $Post_i$ is a dummy variable which takes value 0 if individual $i$ is interviewed in the time period before the invasion and value 1 if the individual is surveyed in the period after the invasion; $\beta$ is the coefficient of interest; $\alpha$ the intercept; and $\varepsilon_i$ an error term. Results estimated with robust standard errors and 95 percent confidence intervals. Estimations include sampling weights. Missing values excluded. Each coefficient corresponds to a separate regression, where Levada total and the different heterogeneity results are estimated using different subsamples in the Levada data between February and March 2022, and Gallup total is estimated using the GWP data between 2021 and 2022. The regression results for Fig. 1 Column 3 are shown in table S3.

### Methods for Fig. 2 in Main Paper

Column 1: Event-study plots showing the monthly effects of the mobilization (mean differences relative to the month before the mobilization) estimated using the following linear regression:

$$y_i = \alpha + \beta D_i + \varepsilon_i$$

where $y_i$ is the survey response of individual $i$ for the relevant sentiment; $D_i$ is a dummy variable which takes value 0 in the month before the mobilization (August 2022) and value 1 in the estimated month; $\beta$ is the coefficient of interest; $\alpha$ the intercept; and $\varepsilon_i$ an error term. (**A–D**) are based on the monthly Levada data between May 2022 and February 2023. Since the migration aspirations question was not asked in Levada during these months, (**E**) is instead based on weekly GWP data between August 24 and October 18, 2022, with the reference period being the week before the mobilization (September 14–20). Results estimated with robust standard errors and 95 percent confidence intervals. Estimations include sampling weights. Missing values excluded.

Each coefficient corresponds to a separate regression. The regression results for Fig. 2 Column 1 are shown in table S4.

Column 2: Bars showing average values before (August 13–September 20) and after (September 21–November 2) the mobilization in the 2022 GWP survey wave, with 95 percent confidence intervals. Mean values weighted using the GWP sampling weights. "DK" and "Refused" responses excluded.

Column 3: Coefficient plots showing the effects of the mobilization estimated using the following linear regression:

$$y_i = \alpha + \beta Post_i + \varepsilon_i$$

where $y_i$ is the survey response of individual $i$ for the relevant sentiment; $Post_i$ is a dummy variable which, for the announcement effect, takes value 0 if individual $i$ is interviewed in the time period before the mobilization and value 1 if the individual is surveyed in the period after the announcement; and, for the completion effect, takes value 0 in the time period after the announcement and value 1 in the period after the completion of the mobilization; $\beta$ is the coefficient of interest; $\alpha$ the intercept; and $\varepsilon_i$ an error term. Results estimated with robust standard errors and 95 percent confidence intervals. Estimations include sampling weights. Missing values excluded. Each coefficient corresponds to a separate regression. Levada total is estimated using the Levada data between August and October 2022. The different heterogeneity results in (A–D) are estimated using different subsamples in the Levada data between August and September 2022 for the announcement effect, and between September and October 2022 for the completion effect, and (E) is estimated using different subsamples in the 2022 GWP data. Young men are the subgroup targeted by the mobilization (men aged 18–27), young women are women in the same age group, and old men and women are aged 28 and above. Gallup total is estimated using the GWP 2022 data, before (August 13–September 20) versus after (September 21–November 2) the mobilization. The regression results for Fig. 2 Column 3 are shown in table S5.

*Methods for Fig. 3 in Main Paper*

(**A**): Event-study plot showing the weekly effect of the Wagner Group rebellion (mean difference relative to the week before the rebellion) estimated using the 2023 GWP survey wave and the following linear regression:

$$y_i = \alpha + \beta D_i + \varepsilon_i$$

where $y_i$ is the survey response of individual $i$ for approval of Putin; $D_i$ is a dummy variable which takes value 0 in the week before the rebellion (June 16–22) and value 1 in the estimated week; $\beta$ is the coefficient of interest; $\alpha$ the intercept; and $\varepsilon_i$ an error term. Results estimated with robust standard errors and 95 percent confidence intervals. Estimations include sampling weights. Missing values excluded. Each coefficient corresponds to a separate regression. The regression results for Fig. 3 (A) are shown in table S6.

(**B**): Coefficient plot showing the effects of the Wagner Group rebellion estimated using the 2023 GWP survey wave and the following linear regression:

$$y_i = \alpha + \beta Post_i + \varepsilon_i$$

where $y_i$ is the survey response of individual $i$ for the relevant sentiment; $Post_i$ is a dummy variable which takes value 0 if individual $i$ is interviewed before (May 23–June 22) the rebellion and value 1 if the individual is surveyed after (June 23–July 29) the rebellion; $\beta$ is the coefficient of interest; $\alpha$ the intercept; and $\varepsilon_i$ an error term. Results estimated with robust standard errors and 95 percent confidence intervals. Estimations include sampling weights. Missing values excluded. Each coefficient corresponds to a separate regression. The regression results for Fig. 3 (B) are shown in table S7.

*Methods for Fig. 4 in Main Paper*

(**A**): Regional correlation between approval of Putin before invasion and effect of invasion on approval of Putin. Data from GWP. Approval of Putin before invasion measured in 2021. Effect of invasion on approval of Putin measured as the difference between the approval of Putin in 2022 and the approval of Putin in 2021. Regional averages include sampling weights. Missing values excluded.

(**B**): Regional correlation between effect of invasion on approval of Putin and casualties per 100,000 inhabitants. Data from GWP and Mediazona (*20*). Effect of invasion on approval of Putin measured as the difference between the approval of Putin in 2022 and the approval of Putin in 2021, in the GWP data. Regional averages include sampling weights. Missing values excluded. Casualties per 100,000 inhabitants measured in Mediazona (and population-adjusted using data from Rosstat) between February 24, 2022, and April 30, 2023.

(**C**): Regional correlation between casualties per 100,000 inhabitants and change in income percentile during war. Data from Mediazona and GWP. Casualties per 100,000 inhabitants measured in Mediazona between February 24, 2022, and April 30, 2023. Change in income percentile during war measured as the difference between the mean per capita income percentile in 2023 and the mean per capita income percentile in 2022, in the GWP data. Regional averages include sampling weights. Missing values excluded.

(**D**): Regional correlation between casualties per 100,000 inhabitants and change in approval of Putin during war. Data from Mediazona and GWP. Casualties per 100,000 inhabitants measured in Mediazona between February 24, 2022, and April 30, 2023. Change in approval of Putin during war measured as the difference between the approval of Putin in 2023 and the approval of Putin in 2022, in the GWP data. Regional averages include sampling weights. Missing values excluded.

The correlation results for Fig. 4 (A–D) are shown in table S8.

*Methods for Fig. 5 in Main Paper*

(**A**): Time-series graph showing the average approval of Putin in the annual GWP survey waves between 2007 and 2023 in Russia, among Russians abroad, and in the rest of the world. Russians abroad include people in other countries who were born in Russia and/or have a Russian nationality. Mean values weighted using the GWP sampling weights. "DK" and "Refused" responses excluded. The results for Fig. 5 (A) are shown in table S9.

(**B**): Bars showing the effect of the invasion on approval of Putin, estimated using the following linear regression:

$$y_i = \alpha + \beta Post_i + \varepsilon_i$$

where $y_i$ is the survey response of individual $i$ for approval of Putin; $Post_i$ is a dummy variable which takes value 0 before the invasion (GWP survey waves 2020–2021) and value 1 after the invasion (GWP survey waves 2022–2023); $\beta$ is the coefficient of interest; $\alpha$ the intercept; and $\varepsilon_i$ an error term. Results estimated with robust standard errors and 95 percent confidence intervals. Estimations include GWP sampling weights. "DK" and "Refused" responses excluded. Each coefficient corresponds to a separate regression. Pro-Putin countries defined as countries in which 50 percent or more approved of Putin in 2020–2021, and anti-Putin countries defined as countries in which less than 50 percent approved of Putin in 2020–2021. The regression results for Fig. 5 (B) are shown in table S10.

## Supplementary Text

Supplementary Results

In this section, we present supplementary results referred to in the main paper.

*Methods for Fig. S1*

(**A**): Bar plot showing the number of interviews per day in the GWP survey waves 2021–2023. Median interview date indicated by 0. Total number of interview days was 62 in 2021, 82 in 2022, and 68 in 2023.

(**B**): Coefficient plot showing a balance test in the Levada data estimated using the following linear regression:

$$y_i = \alpha + \beta Post_i + \varepsilon_i$$

where $y_i$ is the characteristics of individual $i$ for different outcomes which we do not expect to be directly affected by the invasion; $Post_i$ is a dummy variable which takes value 0 if individual $i$ is interviewed in the time period before the invasion (February 2022) and value 1 if the individual is surveyed in the period after the invasion (March 2022); $\beta$ is the coefficient of interest; $\alpha$ the intercept; and $\varepsilon_i$ an error term. Results estimated with robust standard errors and 95 percent confidence intervals. Estimations include sampling weights. Missing values excluded. Each coefficient corresponds to a separate regression. Cut points for income percentiles defined in the period before the invasion (February 2022).

(**C**): Coefficient plot showing a balance test in the Levada data estimated using the following linear regression:

$$y_i = \alpha + \beta Post_i + \varepsilon_i$$

where $y_i$ is the characteristics of individual $i$ for different outcomes which we do not expect to be directly affected by the mobilization; $Post_i$ is a dummy variable which takes value 0 if individual $i$ is interviewed in the time period before the mobilization (August 2022) and value 1 if the individual is surveyed in the period after the mobilization (September 2022); $\beta$ is the coefficient of interest; $\alpha$ the intercept; and $\varepsilon_i$ an error term. Results estimated with robust standard errors and 95 percent confidence intervals. Estimations include sampling weights. Missing values excluded. Each coefficient corresponds to a separate regression.

(**D**): Coefficient plot showing a balance test in the 2022 GWP data estimated using the following linear regression:

$$y_i = \alpha + \beta Post_i + \varepsilon_i$$

where $y_i$ is the characteristics of individual $i$ for different outcomes which we do not expect to be directly affected by the mobilization; $Post_i$ is a dummy variable which takes value 0 if individual $i$ is interviewed in the time period before the mobilization (Aug. 13–Sep. 20) and value 1 if the individual is surveyed in the period after the mobilization (Sep. 21–Nov. 2); $\beta$ is the coefficient of interest; $\alpha$ the intercept; and $\varepsilon_i$ an error term. Results estimated with robust standard errors and 95 percent confidence intervals. Estimations include sampling weights. "DK" and "Refused" responses excluded. Each coefficient corresponds to a separate regression. For "Putin missing" and "Migration missing", the outcomes are dummy variables which take value 1 if the individual has answered "DK" or "Refused" to the approval of Putin or migration aspirations question, respectively, and 0 otherwise.

*Methods for Fig. S2*

Graph showing the mood composition in Russia as measured in the monthly Levada data between January 2021 and April 2023. Shares weighted using the Levada sampling weights. Missing values and "Hard to answer" responses excluded.

*Methods for Fig. S3*

Maps showing the average approval of Putin before the invasion 2006–2021 (**A**) and the effects of the invasion on approval of Putin (**B**) using the GWP data in different countries around the world. Effects of the invasion measured as the difference between the average approval of Putin in 2022–2023 (after the invasion) versus 2020–2021 (before the invasion). Estimations include sampling weights. "DK" and "Refused" responses excluded.

*Methods for Table S11*

In table S11, we present the regression results for a number of sensitivity and robustness analyses for the effect of the invasion on approval of Putin.

The baseline estimation is the same as in the main analysis:

$$y_i = \alpha + \beta Post_i + \varepsilon_i$$

where $y_i$ is the survey response of individual $i$ for the approval of Putin question in the monthly Levada data; $Post_i$ is a dummy variable which takes value 0 if individual $i$ is interviewed in the time period before the invasion (February 2022) and value 1 if the individual is surveyed in the period after the invasion (March 2022); $\beta$ is the coefficient of interest; $\alpha$ the intercept; and $\varepsilon_i$ an error term. Results estimated with robust standard errors and 95 percent confidence intervals, including sampling weights, and excluding missing values.

In the basic specification with control variables, we control for gender and age. In the extended control variables specification, we control for gender, age, marital status, education, income percentile, and geographical area.

In the 6-months specification, we expand the window of analysis and compare the responses of individuals surveyed in December 2021–February 2022 for the pre-invasion period to those surveyed in March–May 2022 for the post-invasion period. For the 1-year specification, we compare responses in the 2021 (May–July) versus 2022 (August–November) survey waves in GWP.

In the Prime Minister and government specifications, we analyze the effects of the invasion on two different survey questions in Levada: "Do you generally approve or disapprove of the activities of the Prime Minister of Russia?" and "Do you generally approve or disapprove of the activities of the government of Russia as a whole?".

In the unweighted specification, we do the baseline estimation but without the sampling weights.

In the missing approve and disapprove specifications, we include missing values and recode them as either approval or disapproval of Putin, respectively.

In the Crimea specifications, we analyze the effects of the 2014 invasion and annexation of Crimea on approval of Putin. For the Crimea invasion effect, we compare survey responses in February versus March 2014, and for the Crimea annexation effect February versus April 2014.

Finally, we do two placebo tests where we analyze the change in approval of Putin in the same months as the invasion (February versus March) but for the years before (2021) and after (2023) the invasion instead.

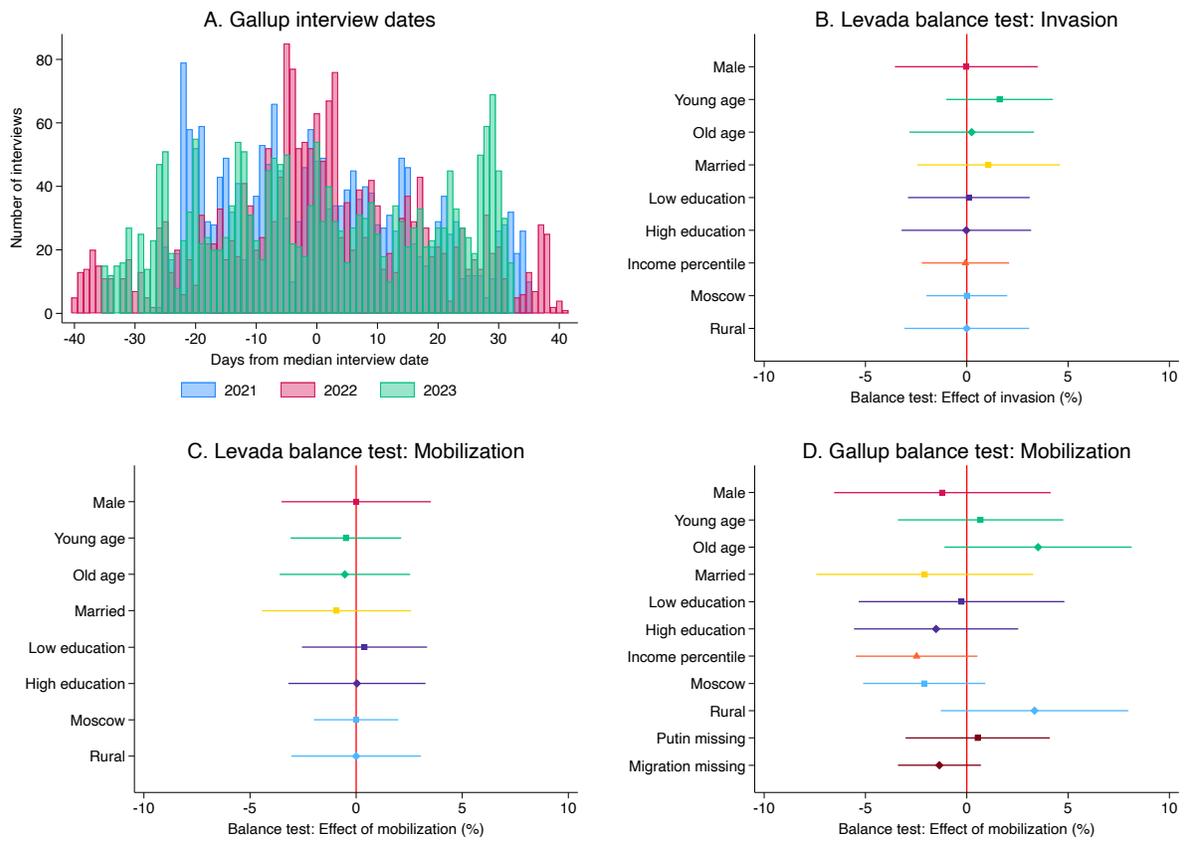

**Fig. S1.**
Survey period lengths and balance tests.

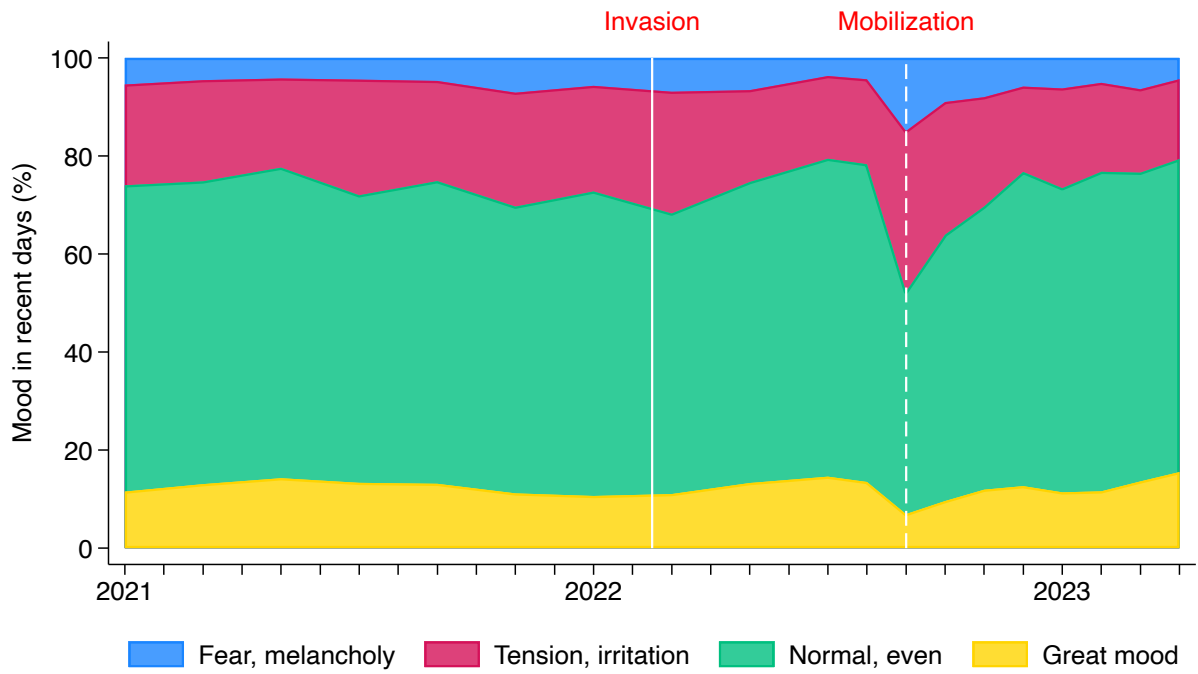

**Fig. S2.**
Mood in the Russian population, January 2021–April 2023.

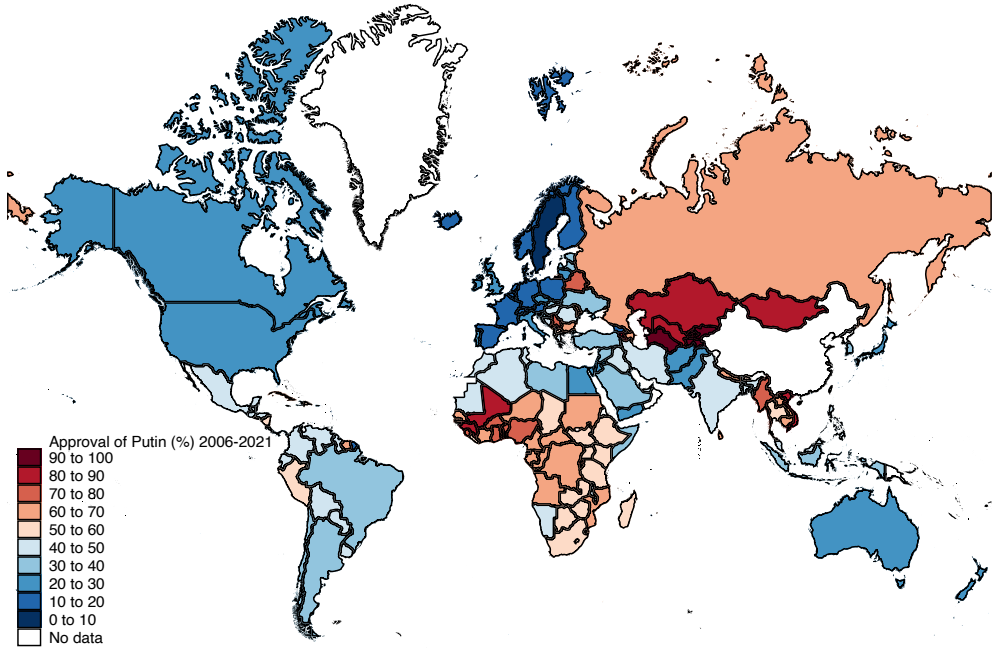

A. Approval of Putin before invasion

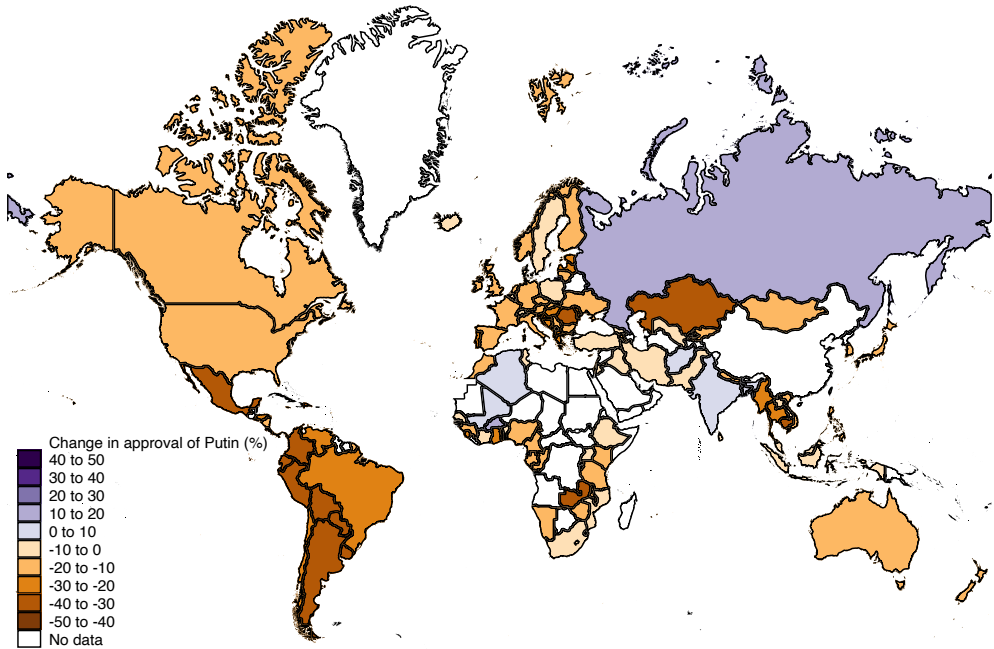

B. Effect of invasion on approval of Putin

**Fig. S3.**
Approval of Putin and invasion effects around the world.

**Table S1.**
GWP sample sizes and survey dates 2007–2023.

| Year | Survey dates Russia | Russia | Russians abroad | Rest of world |
|------|---------------------|--------|-----------------|---------------|
| 2007 | May 1–31 | 2,949 | 1,582 | 100,300 |
| 2008 | May 1–30 | 2,019 | 1,222 | 129,261 |
| 2009 | Apr. 2–Jun. 14 | 2,042 | 1,597 | 134,410 |
| 2010 | Apr. 29–Nov. 8 | 4,000 | 1,150 | 145,706 |
| 2011 | May 8–Jun. 30 | 2,000 | 1,251 | 192,213 |
| 2012 | Feb. 9–Oct. 8 | 3,000 | 1,032 | 226,051 |
| 2013 | Jul. 3–Aug. 8 | 2,000 | 1,134 | 135,539 |
| 2014 | Apr. 22–Jun. 9 | 2,000 | 1,620 | 186,728 |
| 2015 | Jul. 2–Sep. 17 | 2,000 | 1,556 | 144,676 |
| 2016 | Apr. 15–Jun. 22 | 2,000 | 1,489 | 147,235 |
| 2017 | Jun. 9–Aug. 20 | 2,000 | 1,388 | 151,778 |
| 2018 | Jun. 24–Oct. 4 | 2,000 | 1,349 | 149,525 |
| 2019 | Nov. 6–Feb. 10 | 3,003 | 1,273 | 171,977 |
| 2020 | Aug. 19–Oct. 2 | 2,022 | 381 | 127,071 |
| 2021 | May 14–Jul. 14 | 2,001 | 1,059 | 123,842 |
| 2022 | Aug. 13–Nov. 2 | 2,006 | 902 | 140,778 |
| 2023 | May 23–Jul. 29 | 2,017 | 850 | 103,921 |
| Total | | 39,059 | 20,835 | 2,511,011 |

**Table S2.**
Levada sample sizes and survey dates 2021–2023.

| Month | Year | Survey dates | Sample size |
| --- | --- | --- | --- |
| January | 2021 | Jan. 25–31 | 1,616 |
| February | 2021 | Feb. 15–21 | 1,601 |
| March | 2021 | Mar. 22–28 | 1,623 |
| April | 2021 | Apr. 19–25 | 1,614 |
| May | 2021 | May 19–25 | 1,620 |
| June | 2021 | Jun. 14–27 | 3,253 |
| July | 2021 | Jul. 19–25 | 1,619 |
| August | 2021 | Aug. 16–22 | 1,621 |
| September | 2021 | Sep. 20–26 | 1,634 |
| October | 2021 | Oct. 18–24 | 1,636 |
| November | 2021 | Nov. 22–28 | 1,603 |
| December | 2021 | Dec. 13–19 | 1,640 |
| January | 2022 | Jan. 24–30 | 1,626 |
| February | 2022 | Feb. 14–20 | 1,618 |
| March | 2022 | Mar. 27–Apr. 2 | 1,632 |
| April | 2022 | Apr. 27–May 3 | 1,616 |
| May | 2022 | May 2–8 | 1,634 |
| June | 2022 | Jun. 26–Jul. 2 | 1,628 |
| July | 2022 | Jul. 23–29 | 1,617 |
| August | 2022 | Aug. 27–Sep. 2 | 1,612 |
| September | 2022 | Sep. 24–30 | 1,631 |
| October | 2022 | Oct. 23–29 | 1,604 |
| November | 2022 | Nov. 23–29 | 1,601 |
| December | 2022 | Dec. 16–22 | 1,611 |
| January | 2023 | Jan. 25–31 | 1,616 |
| February | 2023 | Feb. 25–Mar. 3 | 1,626 |
| March | 2023 | Mar. 25–31 | 1,633 |
| April | 2023 | Apr. 21–27 | 1,623 |
| Total | | | 47,008 |

**Table S3.**

Regression results for Fig. 1 Column 3 in main text.

| Effect of invasion | Approval of Putin | Optimism about future | Subjective well-being | Positive about West | Migration aspirations |
|---|---|---|---|---|---|
| *Levada* | | | | | |
| Total | 12.57*** | 18.13*** | −0.40 | −19.47*** | −9.28*** |
| | (1.47) | (1.76) | (0.98) | (1.80) | (1.29) |
| | [n = 3,199] | [n = 2,953] | [n = 2,484] | [n = 2,802] | [n = 3,205] |
| *Gender* | | | | | |
| Men | 11.87*** | 18.22*** | 1.95 | −16.09*** | −7.71*** |
| | (2.33) | (2.63) | (1.30) | (2.67) | (2.11) |
| | [n = 1,444] | [n = 1,352] | [n = 1,164] | [n = 1,299] | [n = 1,441] |
| Women | 13.12*** | 18.06*** | −2.69* | −22.40*** | −10.56*** |
| | (1.87) | (2.36) | (1.42) | (2.44) | (1.58) |
| | [n = 1,755] | [n = 1,601] | [n = 1,320] | [n = 1,503] | [n = 1,764] |
| *Age* | | | | | |
| Young | 11.22*** | 10.82** | 0.21 | −25.00*** | −21.16*** |
| | (3.71) | (4.60) | (2.29) | (4.60) | (4.20) |
| | [n = 543] | [n = 490] | [n = 476] | [n = 480] | [n = 536] |
| Middle | 13.96*** | 20.73*** | −1.78 | −19.56*** | −9.62*** |
| | (2.01) | (2.33) | (1.30) | (2.36) | (1.68) |
| | [n = 1,864] | [n = 1,732] | [n = 1,464] | [n = 1,632] | [n = 1,864] |
| Old | 10.06*** | 16.36*** | 2.70 | −15.56*** | −2.41* |
| | (2.52) | (3.07) | (1.87) | (2.93) | (1.27) |
| | [n = 792] | [n = 731] | [n = 544] | [n = 690] | [n = 805] |
| *Marital status* | | | | | |
| Married | 13.04*** | 20.74*** | −0.29 | −18.85*** | −8.96*** |
| | (1.97) | (2.32) | (1.33) | (2.36) | (1.60) |
| | [n = 1,789] | (n = 1,674] | [n = 1,280] | [n = 1,597] | [n = 1,790] |
| Single | 11.96*** | 14.76*** | −0.52 | −20.14*** | −9.57*** |
| | (2.22) | (2.70) | (1.46) | (2.80) | (2.10) |
| | [n = 1,410] | [n = 1,279] | [n = 1,204] | [n = 1,205] | [n = 1,415] |
| *Education* | | | | | |
| Low | 16.51*** | 19.37*** | −0.43 | −17.38*** | −8.54*** |
| | (3.00) | (3.77) | (2.05) | (3.82) | (2.35) |
| | [n = 732] | [n = 663] | [n = 563] | [n = 620] | [n = 736] |
| Middle | 12.60*** | 19.30*** | 0.57 | −16.99*** | −7.91*** |
| | (2.14) | (2.53) | (1.46) | (2.63) | (1.94) |
| | [n = 1,489] | [n = 1,375] | [n = 1,166] | [n = 1,314] | [1,499] |
| High | 9.40*** | 15.31*** | −1.96 | −24.90*** | −12.07*** |
| | (2.76) | (3.21) | (1.71) | (3.23) | (2.45) |
| | [n = 978] | [n = 915] | [n = 755] | [n = 868] | [n = 970] |
| *Income* | | | | | |
| Low | 14.76*** | 18.17*** | 3.53 | −15.07*** | −6.78** |
| | (3.42) | (4.09) | (2.40) | (4.04) | (2.82) |
| | [n = 679] | [n = 610] | [n = 485] | [n = 568] | [n = 685] |
| Middle | 15.36*** | 23.90*** | −2.00 | −22.38 | −9.20*** |
| | (2.16) | (2.56) | (1.33) | (2.57) | (1.82) |
| | [n = 1,454] | [n = 1,332] | [n = 1,129] | [n = 1,283] | [n = 1,461] |
| High | 7.64*** | 11.64*** | 0.24 | −16.59*** | −10.45*** |
| | (2.50) | (2.94) | (1.69) | (3.22) | (2.39) |
| | [n = 1,066] | [n = 1,011] | [n = 870] | [n = 951] | [n = 1,059] |
| *Geographical area* | | | | | |
| Moscow | 2.53 | 4.89 | −1.12 | −7.07 | −7.93 |
| | (5.45) | (5.98) | (2.60) | (5.47) | (4.98) |
| | [n = 280] | [n = 275] | [n = 207] | [n = 272] | [n = 278] |
| Urban | 13.57*** | 20.38*** | −1.05 | −21.21*** | −10.14*** |
| | (1.79) | (2.16) | (1.19) | (2.24) | (1.59) |
| | [n = 2,120] | [n = 1,929] | [n = 1,667] | [n = 1,831] | [n = 2,129] |
| Rural | 13.55*** | 17.28*** | 1.50 | −20.08*** | −7.38*** |
| | (2.92) | (3.49) | (2.13) | (3.60) | (2.41) |
| | [n = 799] | [n = 749] | [n = 610] | [n = 699] | [n = 798] |
| *GWP* | | | | | |
| Total | 25.28*** | 9.69*** | 11.70*** | −25.60*** | −8.93*** |
| | (1.88) | (1.64) | (1.58) | (1.75) | (1.40) |
| | [n = 3,686] | [n = 3,681] | [n = 3,962] | [n = 3,363] | [n = 3,933] |

Note: $*$ $p < 0.1$, $**$ $p < 0.05$, $***$ $p < 0.01$; robust standard errors in parentheses; $n$ = number of observations.

**Table S4.**

Regression results for Fig. 2 Column 1 in main text.

| Effect of mobilization | Approval of Putin | Optimism about future | Subjective well-being | Support for military | | Migration aspirations |
|---|---|---|---|---|---|---|
| *Levada, months* | | | | | *GWP, weeks* | |
| May 2022 | 0.26 (1.32) [n = 3,185] | 1.14 (1.65) [n = 2,940] | –1.39 (0.94) [n = 2,636] | 0.33 (1.45) [n = 3,032] | Aug. 24–30 | 5.06 (4.40) [n = 484] |
| June 2022 | –0.53 (1.33) [n = 3,180] | 1.24 (1.65) [n = 2,947] | | –3.05** (1.48) [n = 3,041] | Aug. 31–Sep. 6 | 3.38 (3.56) [n = 512] |
| July 2022 | 0.53 (1.32) [n = 3,165] | 1.36 (1.65) [n = 2,928] | 1.02 (0.89) [n = 2,651] | –0.52 (1.47) [n = 3,006] | Sep. 7–13 | 2.73 (3.49) [n = 549] |
| August 2022 | . | . | . | . | Sep. 14–20 | . |
| September 2022 | –6.38*** (1.42) [n = 3,183] | –5.24*** (1.73) [n = 2,888] | –11.46*** (1.07) [n = 2,398] | –3.46** (1.50) [n = 3,007] | Sep. 21–27 | 4.45* (2.62) [n = 729] |
| October 2022 | –3.83*** (1.40) [n = 3,145] | –0.99 (1.68) [n = 2,887] | –5.11*** (1.00) [n = 2,484] | –3.09** (1.52) [n = 2,972] | Sep. 28–Oct. 4 | 0.96 (2.95) [n = 567] |
| November 2022 | –3.59** (1.39) [n = 3,132] | –5.22*** (1.73) [n = 2,880] | –3.01*** (0.97) [n = 2,555] | –2.96** (1.50) [n = 2,988] | Oct. 5–11 | 1.82 (2.78) [n = 573] |
| December 2022 | –1.50 (1.35) [n = 3,156] | –1.85 (1.70) [n = 2,869] | –1.41 (0.91) [n = 2,638] | –4.01*** (1.52) [n = 2,979] | Oct. 12–18 | 3.20 (3.55) [n = 495] |
| January 2023 | –0.85 (1.34) [n = 3,152] | –0.94 (1.68) [n = 2,922] | –2.28** (0.92) [n = 2,597] | –1.54 (1.48) [n = 3,012] | | |
| February 2023 | 0.65 (1.31) [n = 3,165] | 2.60 (1.64) [n = 2,921] | –1.55* (0.88) [n = 2,643] | 0.39 (1.44) [n = 3,036] | | |

Note: $*\ p < 0.1$, $**\ p < 0.05$, $***\ p < 0.01$; robust standard errors in parentheses; $n$ = number of observations.

**Table S5.**

Regression results for Fig. 2 Column 3 in main text.

| Effect of mobilization | Approval of Putin | Optimism about future | Subjective well-being | Support for military | Migration aspirations |
|---|---|---|---|---|---|
| *Levada* | | | | | |
| Total | −3.83*** | −0.99 | −5.11*** | −3.09** | |
| | (1.40) | (1.68) | (1.00) | (1.52) | |
| | [n = 3,145] | [n = 2,887] | [n = 2,484] | [n = 2,972] | |
| *Announcement* | | | | | |
| Young men | −10.10* | −6.91 | −10.00*** | −11.32* | |
| | (5.93) | (6.77) | (3.75) | (6.43) | |
| | [n = 258] | [n = 230] | [n = 219] | [n = 244] | |
| Young women | −12.70** | −21.85*** | −10.60*** | −8.75 | |
| | (5.45) | (6.54) | (3.76) | (6.58) | |
| | [n = 259] | [n = 236] | [n = 201] | [n = 232] | |
| Old men | −3.72 | 0.90 | −4.16*** | −1.39 | |
| | (2.35) | (2.78) | (1.57) | (2.25) | |
| | [n = 1,172] | [n = 1,085] | [n = 908] | [n = 1,135] | |
| Old women | −6.89*** | −7.08*** | −18.25*** | −3.32 | |
| | (1.93) | (2.42) | (1.59) | (2.08) | |
| | [n = 1,494] | [n = 1,337] | [n = 1,070] | [n = 1,396] | |
| *Completion* | | | | | |
| Young men | 11.15* | 10.88 | 4.29 | 11.04* | |
| | (5.81) | (6.71) | (3.84) | (6.59) | |
| | [n = 258] | [n = 225] | [n = 210] | [n = 243] | |
| Young women | 5.53 | 18.62*** | 1.47 | 5.19 | |
| | (6.13) | (6.82) | (4.69) | (7.06) | |
| | [n = 244] | [n = 221] | [n = 176] | [n = 214] | |
| Old men | 0.24 | 0.15 | 2.04 | −0.93 | |
| | (2.41) | (2.75) | (1.62) | (2.28) | |
| | [n = 1,161] | [n = 1,058] | [n = 849] | [n = 1,127] | |
| Old women | 2.41 | 3.82 | 11.12*** | −1.03 | |
| | (2.05) | (2.49) | (1.77) | (2.21) | |
| | [n = 1,509] | [n = 1,333] | [n = 1,005] | [n = 1,401] | |
| *GWP* | | | | | |
| Total | 2.14 | −1.63 | −3.42* | 1.34 | 1.81 |
| | (2.44) | (2.14) | (2.07) | (1.93) | (1.68) |
| | [n = 1,740] | [n = 1,753] | [n = 1,973] | [n = 1,892] | [n = 1,939] |
| Young men | | | | | 13.53 |
| | | | | | (8.68) |
| | | | | | [n = 146] |
| Young women | | | | | −4.28 |
| | | | | | (7.70) |
| | | | | | [n = 155] |
| Old men | | | | | 3.26 |
| | | | | | (3.03) |
| | | | | | [n = 514] |
| Old women | | | | | 0.54 |
| | | | | | (1.77) |
| | | | | | [n = 1,066] |

Note: $* \ p < 0.1$, $** \ p < 0.05$, $*** \ p < 0.01$; robust standard errors in parentheses; $n$ = number of observations.

**Table S6.**

Regression results for Fig. 3 (A) in main text.

| Effect of rebellion | Approval of Putin |
|---|---|
| *GWP, weeks* | |
| May 26–<br>Jun. 1 | –0.02<br>(6.04)<br>[n = 401] |
| Jun.<br>2–8 | 5.12<br>(5.30)<br>[n = 459] |
| Jun.<br>9–15 | 7.38<br>(5.18)<br>[n = 467] |
| Jun.<br>16–22 | . |
| Jun.<br>23–29 | 1.39<br>(5.31)<br>[n = 468] |
| Jun. 30–<br>Jul. 6 | –2.22<br>(5.71)<br>[n = 443] |
| Jul.<br>7–13 | 10.56*<br>(5.46)<br>[n = 414] |
| Jul.<br>14–20 | 3.37<br>(5.51)<br>[n = 456] |

Note: $*$ $p < 0.1$, $**$ $p < 0.05$, $***$ $p < 0.01$; robust standard errors in parentheses; $n$ = number of observations.

**Table S7.**

Regression results for Fig. 3 (B) in main text.

| Effect of rebellion | Approval of Putin | Optimism about future | Subjective well-being | Support for military | Positive about West | Migration aspirations |
|---|---|---|---|---|---|---|
| ***GWP*** | | | | | | |
| Total | −0.12 | −0.83 | 0.12 | −2.29 | 0.79 | 1.59 |
| | (2.57) | (2.29) | (2.29) | (2.29) | (1.72) | (1.58) |
| | [n = 1,962] | [n = 1,954] | [n = 2,015] | [n = 2,001] | [n = 1,837] | [n = 2,005] |

Note: $*\ p < 0.1$, $**\ p < 0.05$, $***\ p < 0.01$; robust standard errors in parentheses; $n$ = number of observations.

**Table S8.**

Correlation results for Fig. 4 in main text.

| Pairwise correlation coefficients | Effect of invasion on approval of Putin | Change in income percentile during war | Change in approval of Putin during war |
|---|---|---|---|
| **Approval of Putin before invasion** | –0.87*** [n = 10] | | |
| **Casualties per 100,000 inhabitants** | 0.77*** [n = 10] | 0.75** [n = 10] | –0.15 [n = 10] |

Note: Pearson's correlation coefficients; $*\ p < 0.1$, $**\ p < 0.05$, $***\ p < 0.01$; $n$ = number of observations.

**Table S9.**
Results for Fig. 5 (A) in main text.

| Approval of Putin | Russia | Russians abroad | Rest of world |
|---|---|---|---|
| 2007 | 51% | 87% | 45% |
| 2008 | 75% | 92% | 52% |
| 2009 | 69% | 90% | 49% |
| 2010 | 72% | 80% | 50% |
| 2011 | 64% | 77% | 51% |
| 2012 | 59% | 76% | 44% |
| 2013 | 52% | 78% | 48% |
| 2014 | 81% | 74% | 40% |
| 2015 | 80% | 75% | 42% |
| 2016 | 76% | 78% | 44% |
| 2017 | 71% | 75% | 45% |
| 2018 | 55% | 74% | 47% |
| 2019 | 52% | 72% | 46% |
| 2020 | 56% | | 45% |
| 2021 | 51% | 52% | 45% |
| 2022 | 76% | 31% | 30% |
| 2023 | 70% | 35% | 33% |

**Table S10.**

Regression results for Fig. 5 (B) in main text.

| Effect of invasion | Approval of Putin |
|---|---|
| *GWP* | |
| Russians in Russia | 19.49*** |
| | (1.35) |
| | [n = 7,621] |
| Russians in pro-Putin countries | −8.58** |
| | (4.14) |
| | [n = 801] |
| Russians in anti-Putin countries | −25.63*** |
| | (2.79) |
| | [n = 1,645] |
| Pro-Putin countries | −17.53*** |
| | (0.33) |
| | [n = 150,303] |
| Anti-Putin countries | −14.20*** |
| | (0.24) |
| | [n = 187,801] |

Note: $*$ $p < 0.1$, $**$ $p < 0.05$, $***$ $p < 0.01$; robust standard errors in parentheses; $n$ = number of observations.

**Table S11.**

Regression results for sensitivity and robustness analysis.

| Approval of Putin | Baseline | Controls basic | Controls extended | 6 months | 1 year | Prime Minister | Government |
|---|---|---|---|---|---|---|---|
| Effect of invasion | 12.57*** (1.47) | 12.58*** (1.46) | 12.35*** (1.54) | 14.70*** (0.87) | 25.28*** (1.88) | 11.62*** (1.70) | 15.33*** (1.73) |
| Number of obs. | 3,199 | 3,199 | 2,871 | 9,606 | 3,686 | 3,075 | 3,158 |
| Control variables | No | Few | Many | No | No | No | No |
| Pre-period | Feb. 2022 | Feb. 2022 | Feb. 2022 | Dec. 2021–Feb 2022 | May–Jul. 2021 | Feb. 2022 | Feb. 2022 |
| Post-period | Mar. 2022 | Mar. 2022 | Mar. 2022 | Mar–May 2022 | Aug.–Nov. 2022 | Mar. 2022 | Mar. 2022 |
| Survey question | President | President | President | President | Leadership | Prime Minister | Government |
| Sample weights | Yes | Yes | Yes | Yes | Yes | Yes | Yes |
| Missing values | Excluded | Excluded | Excluded | Excluded | Excluded | Excluded | Excluded |
| Data source | Levada | Levada | Levada | Levada | GWP | Levada | Levada |

| Approval of Putin | Unweighted | Missing approve | Missing disapprove | Crimea invasion | Crimea annexation | Placebo pre | Placebo post |
|---|---|---|---|---|---|---|---|
| Effect of invasion | 12.75*** (1.44) | 12.45*** (1.46) | 12.07*** (1.50) | 8.67*** (1.62) | 13.76*** (1.79) | −1.03 (1.76) | −0.51 (1.30) |
| Number of obs. | 3,199 | 3,250 | 3,250 | 4,716 | 3,163 | 3,163 | 3,168 |
| Control variables | No | No | No | No | No | No | No |
| Pre-period | Feb. 2022 | Feb. 2022 | Feb. 2022 | Feb. 2014 | Feb. 2014 | Feb. 2021 | Feb. 2023 |
| Post-period | Mar. 2022 | Mar. 2022 | Mar. 2022 | Mar. 2014 | Apr. 2014 | Mar. 2021 | Mar. 2023 |
| Survey question | President | President | President | President | President | President | President |
| Sample weights | No | Yes | Yes | Yes | Yes | Yes | Yes |
| Missing values | Excluded | Approve | Disapprove | Excluded | Excluded | Excluded | Excluded |
| Data source | Levada | Levada | Levada | Levada | Levada | Levada | Levada |

Note: $*\ p < 0.1$, $**\ p < 0.05$, $***\ p < 0.01$; robust standard errors in parentheses.